\documentclass[12pt]{article}

\usepackage{graphicx}
\usepackage{epsfig}      

\begin{document}
\begin{center}
{\Large{\bf Hadroproduction of neutral $K^*$-mesons \\ up to LHC energies}} \\ 
\vspace{1.2cm}

{\bf G.H. Arakelyan$^1$, C. Merino$^2$, Yu.M. Shabelski$^{3}$} \\

\vspace{.5cm}
$^1$A.Alikhanyan National Scientific Laboratory \\
(Yerevan Physics Institute)\\
Yerevan, 0036, Armenia\\
e-mail: argev@mail.yerphi.am\\
\vspace{0.1cm}

$^2$Departamento de F\'\i sica de Part\'\i culas, Facultade de F\'\i sica\\
and Instituto Galego de F\'\i sica de Altas Enerx\'\i as (IGFAE)\\
Universidade de Santiago de Compostela\\
15782 Santiago de Compostela\\
Galiza-Spain\\
e-mail: carlos.merino@usc.es\\
\vspace{0.1cm}

$^{3}$Petersburg Nuclear Physics Institute\\
NCR Kurchatov Institute\\
Gatchina, St.Petersburg 188350, Russia\\
e-mail: shabelsk@thd.pnpi.spb.ru
\vskip 0.9 truecm

\vspace{1.2cm}

{\bf Abstract}
\end{center}

We consider the experimental data on neutral $K^*$-meson production on nucleon and nuclear targets. 
The Quark-Gluon String Model quantitatively describes the inclusive density in the midrapidity region, 
as well as the initial energy and A dependences of the produced $K^*$-mesons.

\vskip 1.5cm

\newpage

\vskip -1.5cm
\section{Introduction}

The study of resonance production plays an important role in collisions with nucleon as well as nuclear targets.
 In pp collisions it contributes to the understanding of hadron production, 
as the decay products of resonances represent a large fraction of the final state particles. 

An useful probe of strangeness production is the $K^{*0}(892)$, which is a vector meson with
a mass similar to that of the $\varphi$-meson, but with a strangeness quantum number differing by one unit
of that of the $\varphi$-meson.  

The very short lifetime and the strange valence quark content of the $K^*$-meson 
make the $K^*$-meson production process sensitive to the properties of the dense matter and of strangeness
production, from an early partonic phase. Thus, the measurement of $K^*$-meson properties, such as mass, width, and yields 
can provide significant insight on the dynamics in the dense medium created in heavy-ion collisions.

The hadroproduction of vector $\varphi$-mesons in the frame of Quark-Gluon String Model 
(QGSM)~\cite{KTM,K20} was considered in \cite{amsphi,amsnphi}. 

In this paper we extend the investigation of vector meson production to the case of $K^{*0}$ and 
$\bar K^{*0}$-meson spectra in proton-proton, proton-nucleus, and nucleus-nucleus collisions, for a 
wide range of the initial energy, going up from that of the NA49 experiment, to current the RHIC and LHC energies.

The QGSM is based on the Dual Topological Unitarization (DTU), Regge phenomenology, and nonperturbative notions of QCD, 
and it has been used for the description of secondary particle production at high energies. 
In particular, the QGSM provides quantitative predictions on the inclusive densities of different 
secondaries in the central and beam fragmentation regions in hadron-nucleon~\cite{KTM,K20}, hadron-nucleus 
~\cite{KTMS,Sh1}, and nucleus-nucleus \cite{Sha,AMPSpl} collisions. 

The description of the production of secondary pseudoscalar mesons $\pi$ and $K$,
and of baryons $p$, $\overline{p}$, $\Lambda$, and $\overline{\Lambda}$, was obtained 
in~\cite{KaPiZ,Sh,AMPS,AMPSk}), while vector meson production was considered in~\cite{amsphi,amsnphi,aryer,APSh}. 

The experimental data on neutral $K^{*0}$ and $\bar K^{*0}$-mesons produced by pion and proton beams 
on proton target~\cite{pipkstar,pipkstar2} at not very high energies was already considered in~\cite{AGIres}.  

In the case of collisions on a nuclear target, a new effect was discovered at very high energies, namely the saturation of 
inclusive density of secondaries~\cite{CKTr,amshsv,ACKS}. This saturation effect is also sucsessfully described by 
QGSM~\cite{ACKS,MPSppb,MPSpa}.

\section{Meson inclusive spectra in the QGSM}

In order to produce quantitative results for the inclusive spectra of secondary hadrons,
a model for multiparticle production is needed. It is for that purpose that we have
used the QGSM~\cite{KTM,K20} in the numerical calculations presented below.

\begin{figure}[htb]
\vskip -6.5cm
\hskip 3.5cm
\includegraphics[width=0.7\hsize]{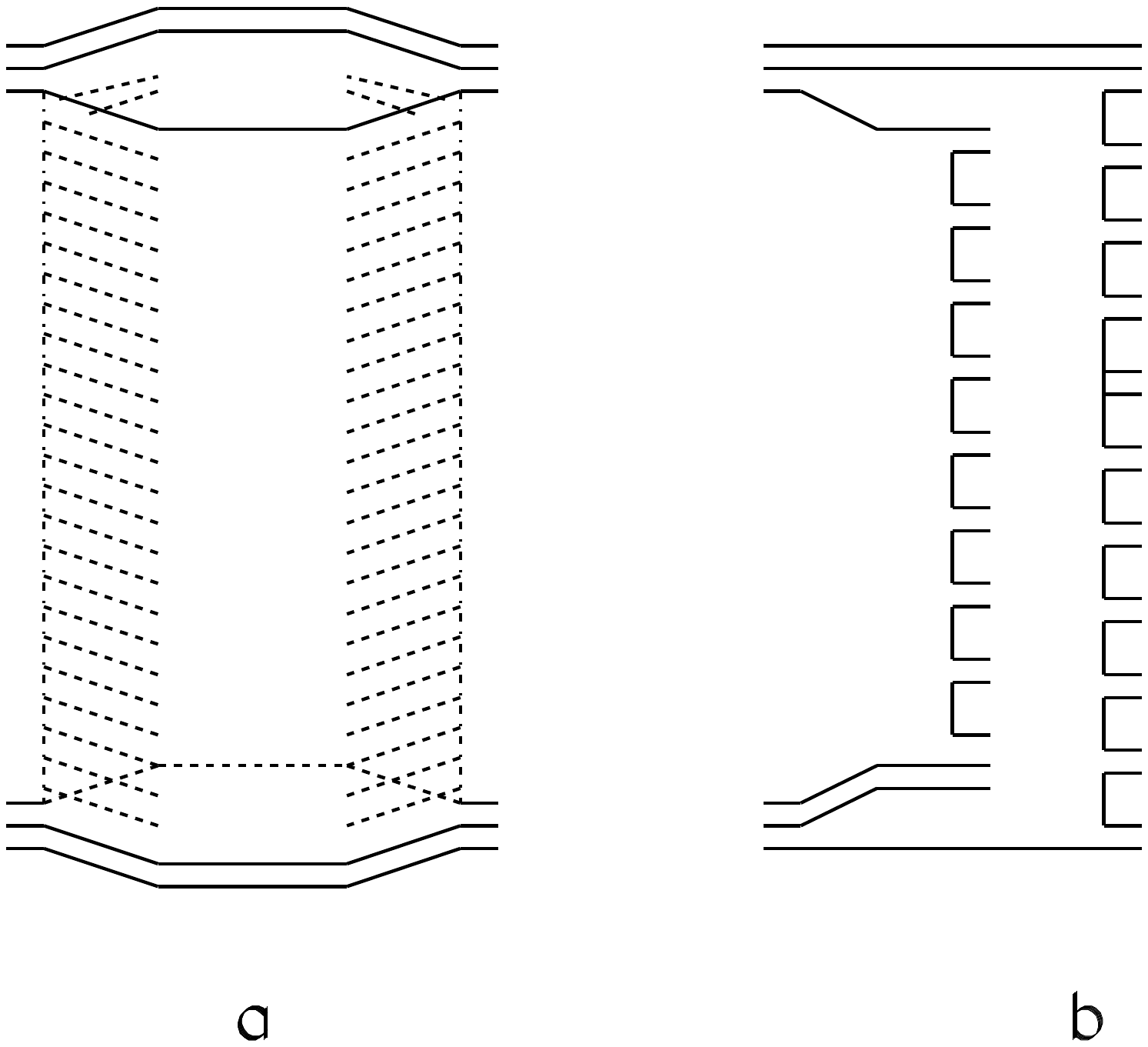}
\vskip -8.0cm
\hskip 1.75cm
\includegraphics[width=0.8\hsize]{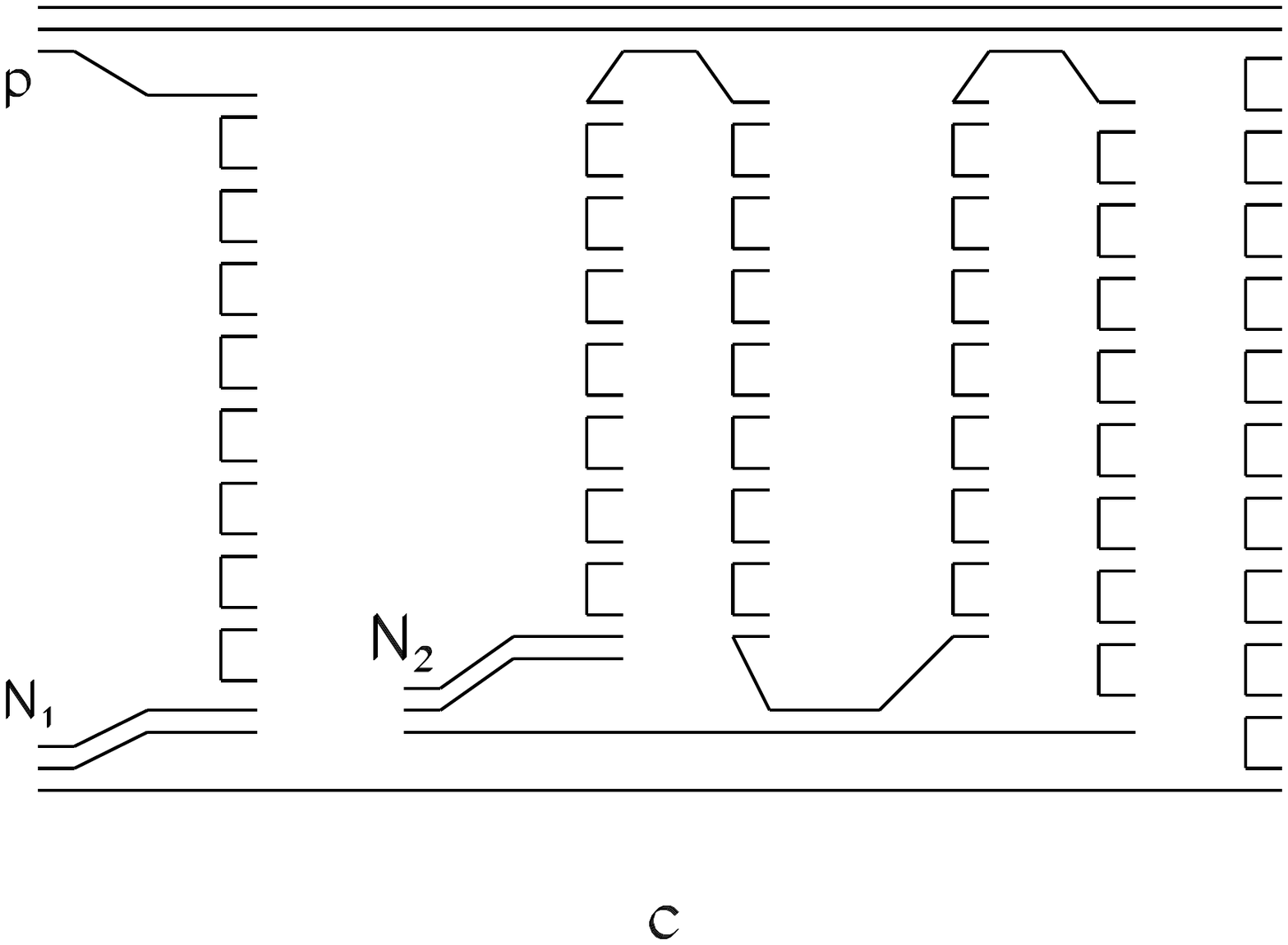}
\vskip -1.25cm
\caption{\footnotesize
(a) Cylindrical diagram representing the Pomeron exchange
within the Dual Topological Unitarization (DTU) classification
(quarks are shown by solid lines);
(b) Cut of the cylindrical diagram corresponding to the single-Pomeron
exchange contribution in inelastic $pp$ scattering;
(c) Diagram corresponding to the inelastic interaction
of an incident proton with two target nucleons $N_1$ and $N_2$
in a $pA$ collision.}
\label{fig:fi1}
\end{figure}

In the QGSM, the high energy hadron-nucleon, hadron-nucleus, and nucleus-nucleus interactions are treated 
as proceeding via the exchange of one or several Pomerons, and all elastic and inelastic processes result 
from cutting through or between Pomerons~\cite{AGK}. Each Pomeron corresponds to a cylinder diagram (see
Fig.~1a), in which the cylinder boundaries are drawn by the dash-dotted vertical lines. The surface of 
the cylinder is schematically depicted by dashed lines, while the solid lines at the top and bottom of 
the cylinder represent, respectively, the beam and the target quarks, which interaction is mediated by 
the Pomeron exchange.

The cut through the cylinder produces two showers of secondaries, i.e. quark-antiquark pairs shown in 
Fig~1b by solid lines. The inclusive spectrum of secondaries is then determined by the convolution of diquark, 
valence quark, and sea quark distributions in the incident particles, $u(x,n)$, with the fragmentation 
functions of quarks and diquarks into the secondary hadrons, $G(z)$. Both functions $u(x,n)$ and $G(z)$ are 
determined by the appropriate Reggeon diagrams~\cite{Kai}.

Note that the quark (antiquark) distributions $u(x,n)\ (\bar{u(x,n)})$ differ from the standard PDF's extracted from fits to 
experimental data because theoretically they are taken to be valid at the rather low $Q^2$ relevant for soft processes,
while the PDF distributions are obtained by fixing the behaviour at large $Q^2$.
The diquark and quark distribution functions depend on the number $n$ of cut Pomerons in the considered diagram. 
In the following calculations we have used the prescription given in reference~\cite{KTMS}.

For a nucleon target, the inclusive rapidity, $y$, or Feynman-$x$, $x_F$, spectrum of a secondary hadron $h$ 
has the form~\cite{KTM,K20}:
\begin{equation}
\frac{dn}{dy}\ = \
\frac{x_E}{\sigma_{inel}}\cdot \frac{d\sigma}{dx_F}\ = 
\sum_{k=1}^\infty w_k\cdot\phi_k^h (x) \ ,
\end{equation}
where $x_{E} = E/E_{max}$, is the relative energy of the secondary particle, the functions $\phi_{k}^{h}(x)$ determine 
the contribution of diagrams with $k$ cut Pomerons, and $w_k$ is the relative weight of this diagram, determined as 
\begin{equation}
w_k= \sigma_k/(\sigma_{tot} - \sigma_{el})\ \;,
\end{equation}

Here, for the production of $K^*$-mesons, we neglect by the contribution of diffraction and disssociation processes.

In the case of $pp$ collisions:
\begin{eqnarray}
\phi_k^{h}(x) &=& f_{qq}^{h}(x_{+},k) \cdot f_{q}^{h}(x_{-},k) +
f_{q}^{h}(x_{+},k) \cdot f_{qq}^{h}(x_{-},k) \nonumber\\
&+& 2(k-1)\cdot f_{s}^{h}(x_{+},k) \cdot f_{s}^{h}(x_{-},k)\ \  , \\
x_{\pm} &=& \frac{1}{2}[\sqrt{4m_{T}^{2}/s+x^{2}}\pm{x}]\ \ , 
\end{eqnarray}
where $m_T = \sqrt{m^2 + p^2_T}$ is the transverse mass of the produced hadrons, and $f_{qq}$, $f_{q}$, and $f_{s}$ 
correspond to the contributions of diquarks, valence quarks, and sea quarks, respectively. The contribution 
of sea quarks and sea antiquarks are assumed to be equal, the difference between quarks and antiquarks being
the valence quarks contribution~\cite{K20,Kai}.

These contributions are determined by the convolution of the diquark and quark distributions with the 
fragmentation functions, e.g.,
\begin{equation}
f_i^h(x_{\pm},k)\ =\ \int\limits_{x_{\pm}}^1u_i(x_1,k)G_i^h(x_{\pm}/x_1) dx_1\ \;,
\end{equation}
where i= qq-diquarks, valence $q, \bar q,$  and sea quarks.

In this paper we have used the distribution functions in the colliding particles of valence and sea quarks, and of diquarks, 
obtained in~\cite{K20,amsphi,Sh,KaPi}. 

For the case of meson production, one has the diagram corresponding to quark and diquark 
fragmentation to secondary mesons shown in Fig.~2:
\begin{figure}[htb]
\label{meson}
\vskip -11.5cm
\hskip 2.0cm
\includegraphics[width=1.0\hsize]{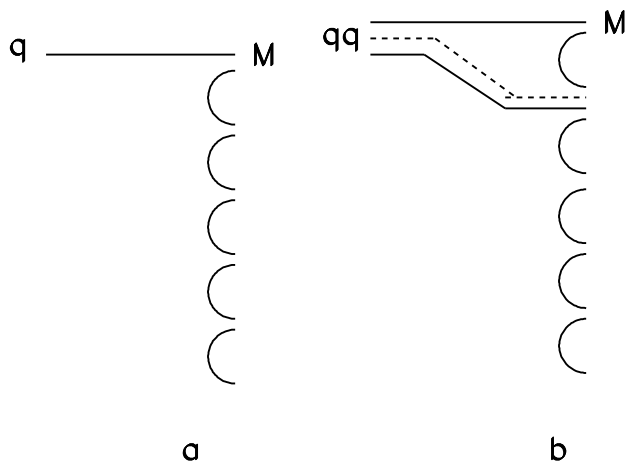}
\vskip -2.5cm
\caption{QGSM diagrams corresponding to fragmentation of (a) quark and 
(b) diquark into a secondary meson $M$.}
\label{fig:fig2}
\end{figure}
 
As for the fragmentation function of quarks and diquarks into neutral vector mesons $K^{*0}$ 
and $\bar K^{*0}$, $G^{K^{*}}_{q,qq}(z)$, we have used the corresponding fragmentation functions for 
pseudoscalar $K$-meson production given in ref.~\cite{Sh,KaPi}. These fragmentation functions were 
obtained using the Reggeon counting rules and the simplest extrpolation~\cite{K20,Kai}. 
The question of spin dependence of fragmentation functions was analyzed in the paper~\cite{AGIres}.
Following this paper, we assume that the functional form of the fragmentation functions is spin independent, 
the spin dependence only appearing in the value of the normalisation parameter $a^{K^*}$ (see below in this section).

With these assumptions, the quark/antiquark fragmentation functions $G^{K^{*}}_{q}(z)$ 
can be written as:

\begin{eqnarray}
G^{K^{*0}}_{d}(z) &=& a^{K^{*}}(1-z)^{-\alpha_{\varphi}(0)+\lambda}(1+b^{K}_{1}z)\;,\nonumber \\
G^{\bar{K}^{*0}}_{d}(z) &=& G^{\bar{K}^{*0}}_{u}(z) 
=a^{K^{*}}(1-z)^{-\alpha_{\varphi}(0)+\lambda+1} \;,\nonumber \\
G^{K^{*0}}_{\bar{s}}(z) &=& 
G^{\bar{K}^{*0}}_{s}(z) = bz^{1-\alpha_{\varphi}(0)}
(1-z)^{-\alpha_{R}(0)+\lambda}+ \\
&+& a^{K^{*}}(1-z)^{-\alpha_{R}(0)+\lambda+2(1-\alpha_{\varphi}(0))}\;, \nonumber \\
G^{\bar{K}^{*0}}_{\bar{s}}(z) &=& G^{K^{*0}}_{s}(z) =
a^{K^{*}}(1-z)^{-\alpha_{R}(0)+\lambda+2(1-\alpha_{\varphi}(0))}\;.\nonumber
\label{eq:eq6}
\end{eqnarray}
Correspondingly, the fragmentation functions for diquarks, $G^{K^{*}}_{qq}(z)$, are:
\begin{eqnarray}
G^{\bar{K}^{*0}}_{uu}(z) &=& G^{K^{*0}}_{uu}(z) =
a^{K^{*}}(1-z)^{-\alpha_{\varphi}(0)-2\alpha_{N}(0)+\lambda+2}\;, \nonumber\\
G^{\bar{K}^{*0}}_{ud}(z) &=&
a^{K^{*}}(1-z)^{-\alpha_{\varphi}(0)-2\alpha_{N}(0)+\lambda+2}(1-z/2)\;, \\
G^{K^{*0}}_{ud}(z) &=&
a^{K^{*}}(1-z)^{-\alpha_{\varphi}(0)-2\alpha_{N}(0)+\lambda+2}
(1+b^{K}_{2} z/2)\;, \nonumber
\label{eq:eq7}
\end{eqnarray}
where $\alpha_{\varphi}(0) \approx 0$, $b^{K}_{1} \approx 2$, $b^{K}_2 \approx 5$, 
and $b \approx 0.4$.

The parameter $\lambda=2 \alpha^\prime_R <p_\bot^2>_{K^*}$, with $\alpha^\prime_R \approx 1$ is 
the slope of the vector Regge trajectory, and $<p_\bot^2>_{K^*}$ is the average transverse squared momenta of 
the produced meson.

In ref.~\cite{AGIres}, the relations for the probabilities of the production of the light 
and strange pseudoscalar and vector mesons were obtained, by using the predictions of the
resonance decay model~\cite{18}. 

The normalization parameter $a_K^*$ in the fragmentation functions of eqs. (6) and (7) is related to 
parameter $a_K$, earlier determined in QGSM for the description of $K$-meson production~\cite{KaPiZ,Sh},
by the equation:
\begin{equation}
\label{11}
(a^{K^*}/a^K)^2=\frac{<k^2_{\bot}>_{K}}{4m^2_q} \;,
\end{equation}
where $m_q=0.415\pm0.015$ is the transverse mass of the constituent quark \cite{18}. 
Taking into account that $<{k^2_{\bot}}>_K \approx 0.21$ GeV$^2$, and $a_K \approx 0.27 $ \cite{KaPiZ,Sh}, 
we find 
\begin{eqnarray}
\label{12}
a^{K^*} \approx 0.15\;.
\end{eqnarray}

In the calculation of the inclusive spectra of secondaries produced
in $pA$ collisions we should consider the possibility of one
or several Pomeron cuts in each of the $\nu$ blobs of the proton-nucleon
inelastic interactions.
For example, in Fig.~1c it is shown one of the diagrams contributing
to the inelastic interaction of a beam proton with two nucleons from the target.
In the blob of the proton-nucleon$_1$ interaction one Pomeron is cut,
and in the blob of the proton-nucleon$_2$ interaction two Pomerons
are cut. 

The contribution of the diagram in Fig.~1c to the inclusive spectrum is presented in refs.~\cite{amsphi,Sh1,Shab}.

The total number of exchanged Pomerons becomes as large as
\begin{equation}
\langle k \rangle_{pA} \sim
\langle \nu \rangle_{pA} \cdot \langle k \rangle_{pN} \;,
\end{equation}
where $\langle \nu \rangle_{pA}$ is the average number of inelastic collisions inside the nucleus
(about 4 for heavy nuclei at fixed target energies).

The process shown in Fig.~1c satisfies~\cite{Sh3,BT,Weis,Jar} the condition
that the absorptive parts of the hadron-nucleus amplitude are determined by
the combination of the absorptive parts of the hadron-nucleon amplitudes.

In the case of a nucleus-nucleus collision, in the projectile fragmentation region
we use the approach~\cite{Sha,Shab,JDDS}, with the beam of
independent nucleons of the projectile interacting with the target nucleus,
what corresponds to the rigid target approximation~\cite{Alk} of the
Glauber Theory. In the target fragmentation region, on the contrary, the
beam of independent target nucleons interact with the projectile nucleus,
the two approaches coinciding in the central region. The corrections due to energy
conservation play here a very important role when the initial energy is not
very high. 

\section{Neutral $K^*$ mesons production in pp collisions}

In Fig. 3 we compare the experimental data of NA49 Collabortion~\cite{na49kstar} on the rapididty dependence 
of $dn/dy$ density of $K^{*0}$ and ${\bar K}^{*0}$ production in $pp$ collisions at 158~GeV/c with the results of the
QGSM calculations. The full dots and triangles represent 
the measured experimental data on $K^{*0}$ and ${\bar K}^{*0}$, respectively.  
\begin{figure}[htb]
\vskip -12.25cm
\hskip 2.0cm
\includegraphics[width=1.2\hsize]{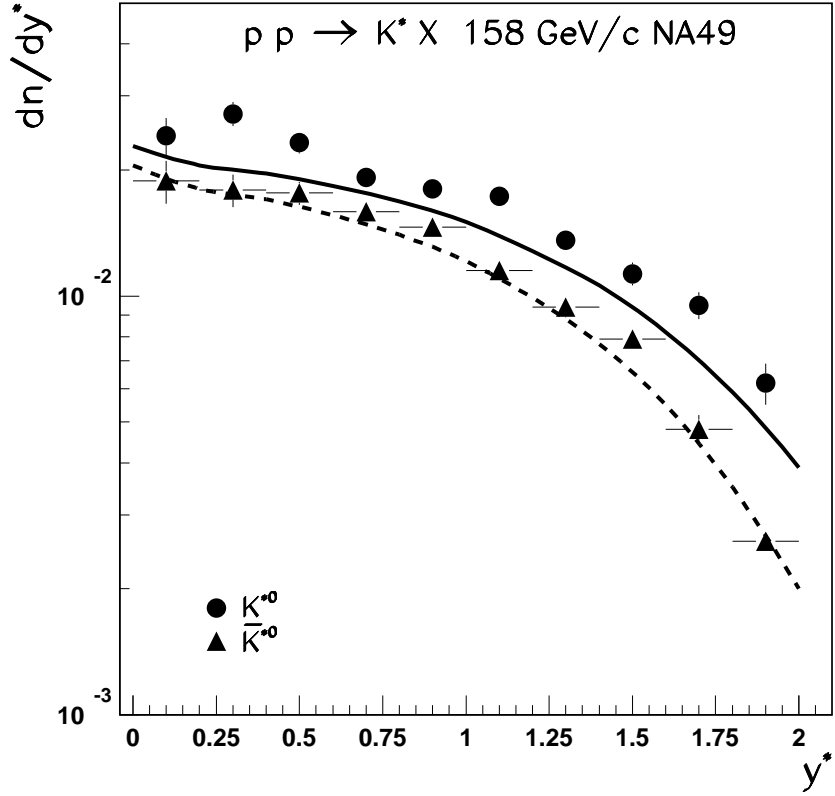}
\vskip -0.75cm
\caption{\footnotesize
Comparison of the results of the QGSM calculations for the $y$-dependence $dn/dy$ of $K^{*0}$ (full line) 
and $\bar K^{*0}$ (dashed line) mesons produced in $pp$ collisions at 158~GeV/c, with 
the corresponding experimental data by the NA49 Collaboration~\cite{na49kstar}.}
\label{fig·fig3}
\end{figure}

Here the theoretical curves are only shown for the rapidity region $y^*\geq$~0, where the experimental data were 
measured, and we omit them in the negative $y^*$ region where NA49 presents the mirror reflection data. 
  
As it can be seen in Fig.~3, the agreement between the theoretical curve and the experimental data is rather good
for $\bar K^{*0}$-meson production in the whole experimental region, while For $K^{*0}$-meson production 
the theoretical curve is slightly lower than the experimental points.

\begin{figure}[htb]
\vskip -12.0cm
\hskip 2.0cm
\includegraphics[width=1.2\hsize]{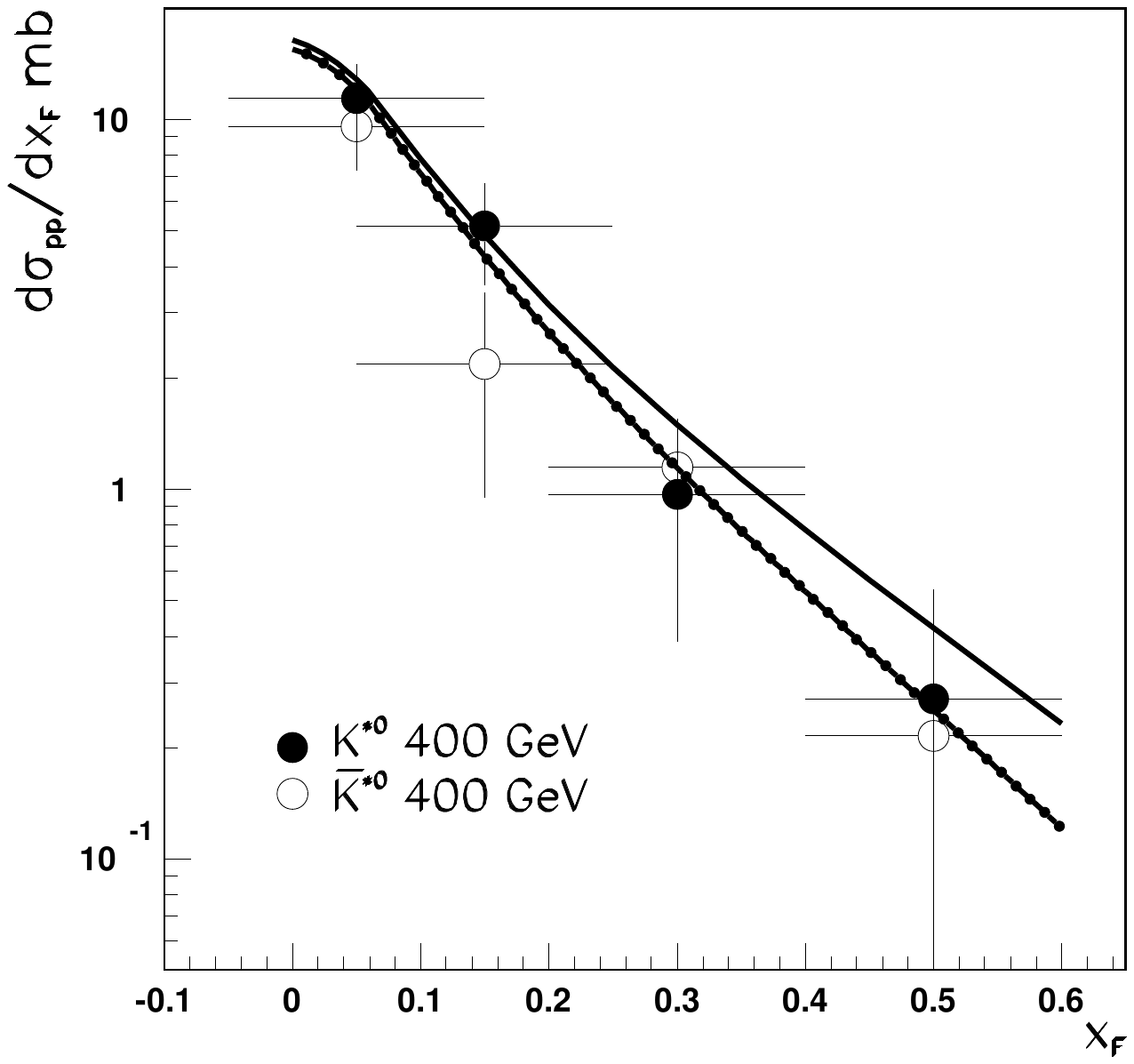}
\vskip -0.75cm
\caption{\footnotesize
Comparison of the results of the QGSM calculations for the $x_F$-spectra $d\sigma/dx_F$ of $K^{*0}$ (full curve)
and $\bar K^{*0}$ (dashed curve) mesons produced in $pp$ collisions at 400~GeV/c, with the corresponding 
experimental data by the LEBC-EHS Collaboration~\cite{agben}.}
\label{fig:fig4}
\end{figure}

In Fig.~4, we compare the results of the QGSM calculations for the $x_F$-spectra $d\sigma/dx_F$ 
of $K^{*0}$ (full curve) and $\bar K^{*0}$ (dashed curve) mesons produced in $pp$ collisions at 400~GeV/c, with
the corresponding experimental data by the LEBC-EHS Collaboration~\cite{agben}. 
The comparison of the theoretical curves with the experimental data provides a reasonable agreement for the whole measured $x_F$ range.
\begin{figure}[htb]
\label{kstars}
\vskip -12.0cm
\hskip 2.0cm
\includegraphics[width=1.2\hsize]{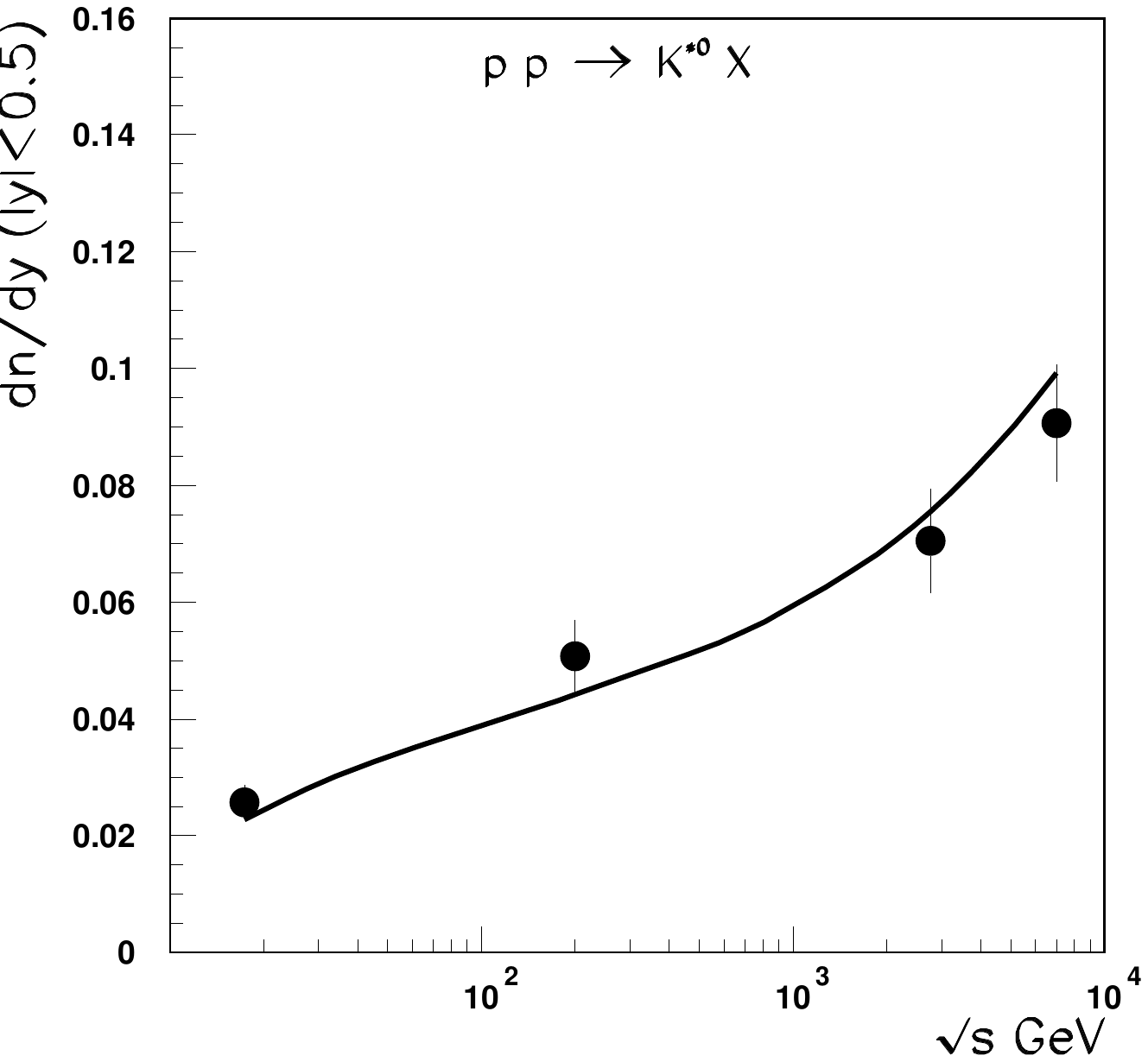}
\vskip -0.75cm
\caption{\footnotesize
Comparison of the experimental data~\cite{na49kstar,star200kstar,alicepp7} on the $\sqrt{s}$-dependence of the 
density $dn/dy (|y|<0.5)$ of average $(K^{*0}+{\bar K}^{*0})/2$ mesons 
produced in $pp$ collisions, with the results of the corresponding QGSM calculations.}
\label{fig:fig5}
\end{figure}

The energy dependence of the density $dn/dy$ $(y = 0)$ of neutral strange mesons produced in $pp$ collisions is shown 
in Fig.~5. Since the experimental data from RHIC and those by the ALICE Collaboration were presented for average
$(K^{*0}+{\bar K}^{*0})/2$ mesons, here we also plot the NA49 Collaboration data for the average $K^{*0}$ meson. The results of the
model calculation show to be in agreement with the experimantal data.

In the Table~1 we present the experimental data on the densities of neutral $K^*$-mesons produced in $pp$ collisions 
for enery region ranging from the NA49 Collaboration to LHC. Note that the NA49 experiment separately measured $K^{*0}$ and 
${\bar K}^{*0}$ mesons, while at RHIC and the ALICE Collaboration at LHC the average density of neutral $K^*$ mesons 
($K^{*0}$ + ${\bar K}^{*0}$)/2 was measured. 

\begin{center}
\vskip -.4cm
\begin{tabular}{|c|c|c|c|c|} \hline
     &    &  &   & \\
Reaction & Produced & Energy & Experimental data & QGSM \\
	& particle  & $\sqrt{s}\ (GeV)$ & $dn/dy\ |y|\leq 0.5$ & \\
\hline
     &    &  &   & \\
$p + p$ & $K^{*0}$ & 17.3 & 0.0257 $\pm$ 0.0031 $\pm$ 0.0023~\cite{na49kstar} & 0.0228  \\ \hline
   &    &  &   & \\
$p + p$ & ${\bar K}^{*0} $ & 17.3 & 0.0183 $\pm$ 0.0027 $\pm$ 0.0016~\cite{na49kstar} & 0.0205 \\ \hline
   &    &  &   &\\
	$p + p$ & $(K^{*0}+{\bar K}^{*0})/2$ & 200.0 & 0.0508 $\pm$ 0.0017 $\pm$ 0.0061~\cite{star200kstar} & 0.0443 \\ \hline
   &    &  &   &\\
$p + p$ & $(K^{*0}+{\bar K}^{*0})/2$ & 2760.0 & 0.0705 $\pm$ 0.0007 $\pm$ 0.009~\cite{alicepppb2017} & 0.0755 \\ \hline
   &    &  &   &\\
$p + p$ & $(K^{*0}+{\bar K}^{*0})/2$ & 7000.0 & 0.097 $\pm$ 0.00004 $\pm$ 0.01~\cite{alicepp7} & 0.106 \\ \hline

\end{tabular}
\end{center}
\vskip -0.1cm
Table~1: {\footnotesize Experimental data on $dn/dy$, $|y|\leq 0.5$, 
of neutral $K^*$-mesons production in $pp$ collisions at different energies, together with the results
of the corresponding QGSM calculations.} 
 
\section{$K^*$-meson production on nuclear targets at not very high energies}

In this section we consider $K^{*0}$ and $\bar K^{*0}$ meson production in proton and nucleus 
collisions on nuclear targets at the energies of NA49, HERAb, and RHIC. 

In Fig. 6 we compare the results of the QGSM calculations for the rapidity spectra $dn/dy$ of $K^{*0}$ (full line) and 
${\bar K}^{*0}$ (dashed line) mesons produced in Pb+Pb collisions at 158~GeV/c, with the corresponding
experimental data by the NA49 Collaboration~\cite{na49kstar}. As it can be seen in the figure, the agreement is reasonable
for the case ${\bar K}^{*0}$ production, while for $K^{*0}$ production the theoretical curve is slightly lower than experiment.

The experimental data on the inclusive cross section rapidity distribution, $d\sigma/dy^*$, for $K^{*0}$ and 
${\bar K}^{*0}$ meson production in proton collisions with different nuclei, measured by the HERAb 
Collaboration~\cite{HERAb} at $\sqrt s = 41.6$~Gev are compared in Fig.~7 with the results of the QGSM calculations. 
The agreement appears to be rather good.

In Fig.~8 we compare the experimental data by the HERAb Collaboration on the atomic number A dependence of $K^{*0}$ and 
${\bar K}^{*0}$ meson production in $pA$ collisions at $\sqrt{s} = 41.6$~Gev~\cite{HERAb}, with the corresponding
QGSM predictions. The solid curve shows the $K^{*0}$-meson production. The curve for ${\bar K}^{*0}$-mesons production
practically coincides with the curve for $K^{*0}$, and so it is not shown.

In Table 2 we compare the corresponding results of QGSM calculations 
with the experimental data by the HERAb Collaboration~\cite{HERAb} on
the cross section $\sigma_{vis}$ $(mb)$

\newpage
\begin{figure}[htb]
\label{pkst41}
\vskip -11.25cm
\hskip 3.0cm
\includegraphics[width=0.95\hsize]{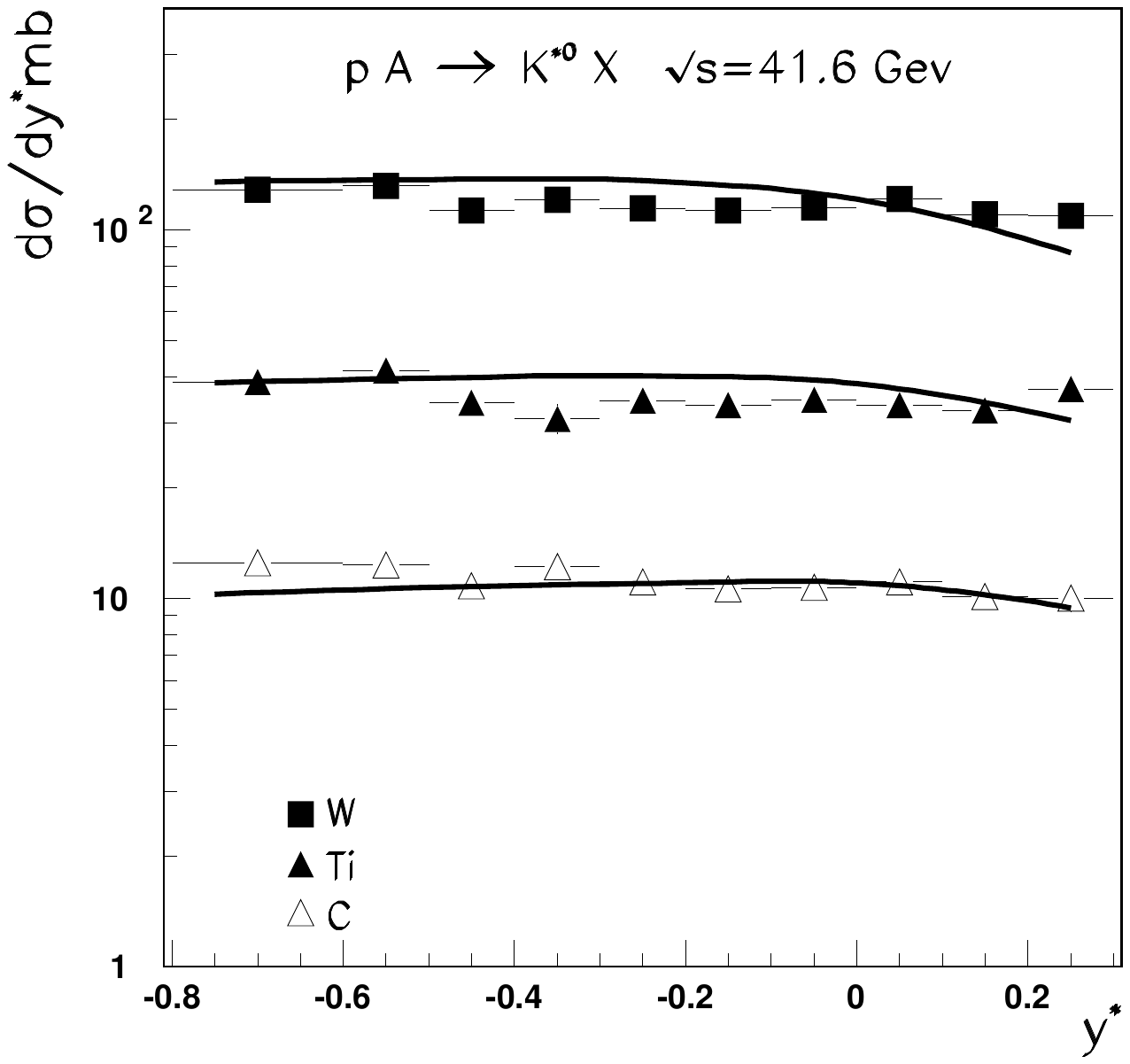}
\vskip -10.5cm
\hskip 3.0cm
\includegraphics[width=0.95\hsize]{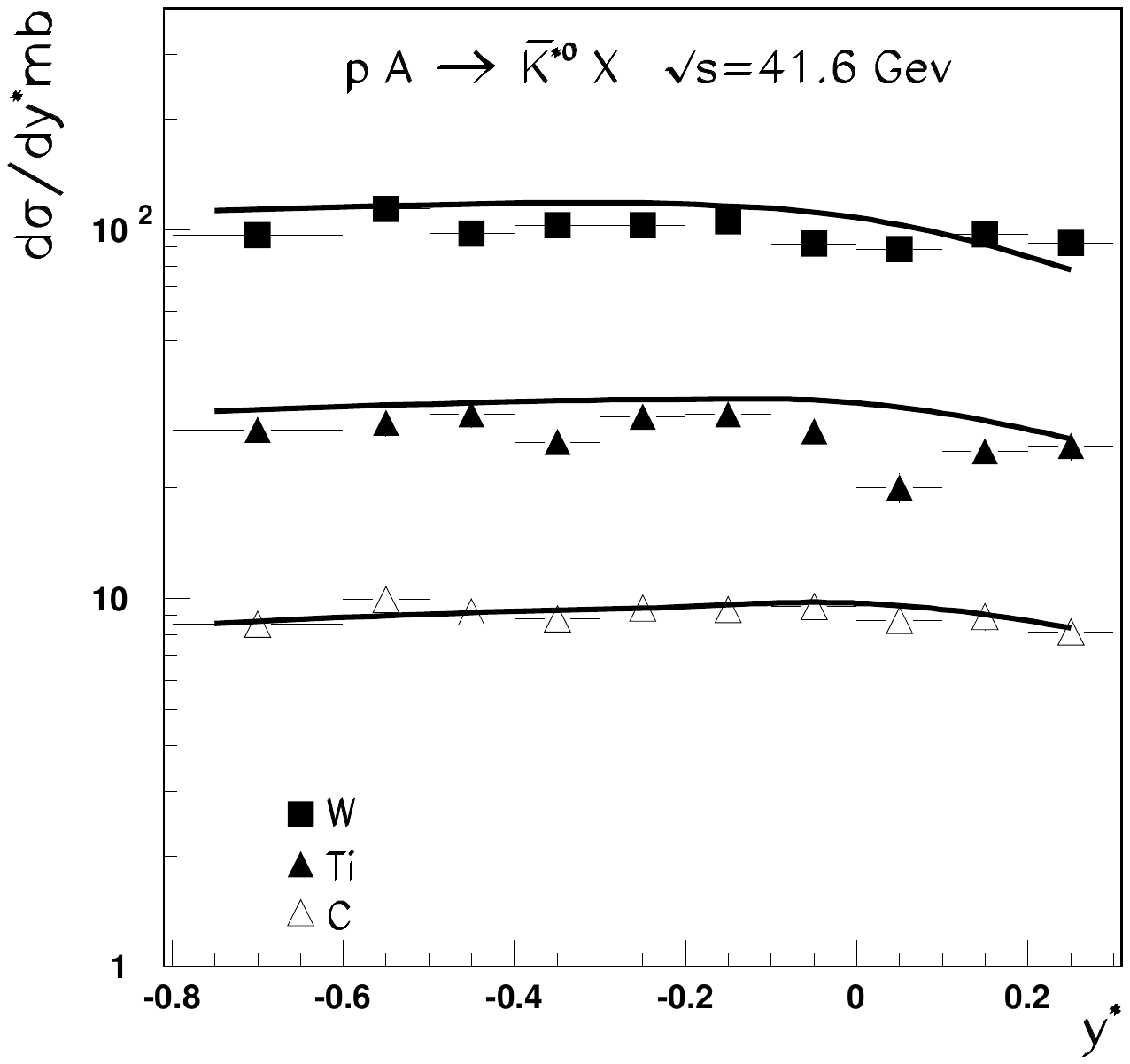}
\vskip -1.0cm
\caption{\footnotesize
Comparison of the results of QGSM calculations for the $y$-spectra $d\sigma/dy$ of 
$K^{*0}$ (upper panel) and ${\bar K}^{*0}$ (lower panel) mesons produced in proton-nucleus 
collisions on different nuclear targets at $\sqrt s = 41.6$~GeV, with the corresponding
experimental data for nuclear targets $C$, $Ti$, and $W$~\cite{HERAb}.}
\label{fig:fig6}
\end{figure}

\noindent
measured in a rather short rapidity region $-0.75 \leq y \leq 0.25$, and with the
results of the extrapolation of data on $K^{*0}$ and ${\bar K}^{*0}$ meson production cross section $\sigma_{prod}$ $(mb)$
for the whole rapidity region. 
The theoretical results for the cross section $\sigma_{prod}$ are lower than the experimental data. This can be connected 
with a different behaviour of the theoretical curves and the experimental extrapolations in the whole rapidity region. 
On the other hand, the theoretical results for $\sigma_{vis}$ consistently coincide with the experimental measurements. 
\newpage

\begin{figure}[htb]
\vskip -13.25cm
\hskip 1.0cm
\includegraphics[width=1.2\hsize]{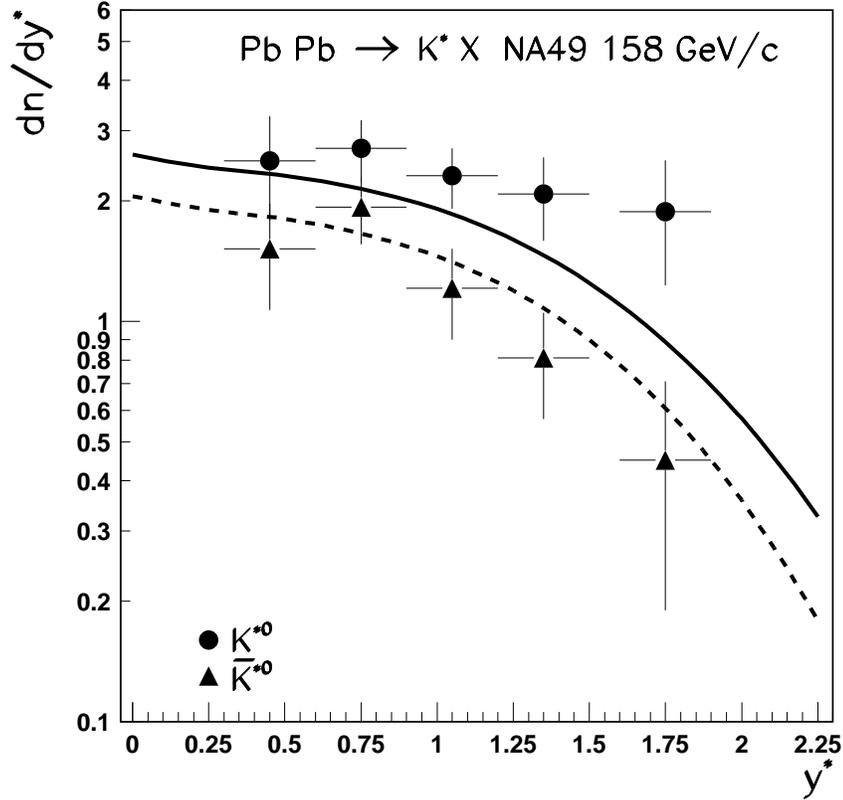}
\vskip -0.75cm
\caption{\footnotesize
Comparison of the results of the QGSM calculations for the rapidity spectra $dn/dy$ of 
$K^{*0}$ (full line) and ${\bar K}^{*0}$ (dashed line) mesons produced 
in Pb+Pb collisions at 158~GeV/c, with the corresponding experimental data by the
NA49 Collaboration~\cite{na49kstar}.}
\label{fig:fig7}
\end{figure}

\begin{figure}[htb]
\vskip -13.5cm
\hskip 1.0cm
\includegraphics[width=1.2\hsize]{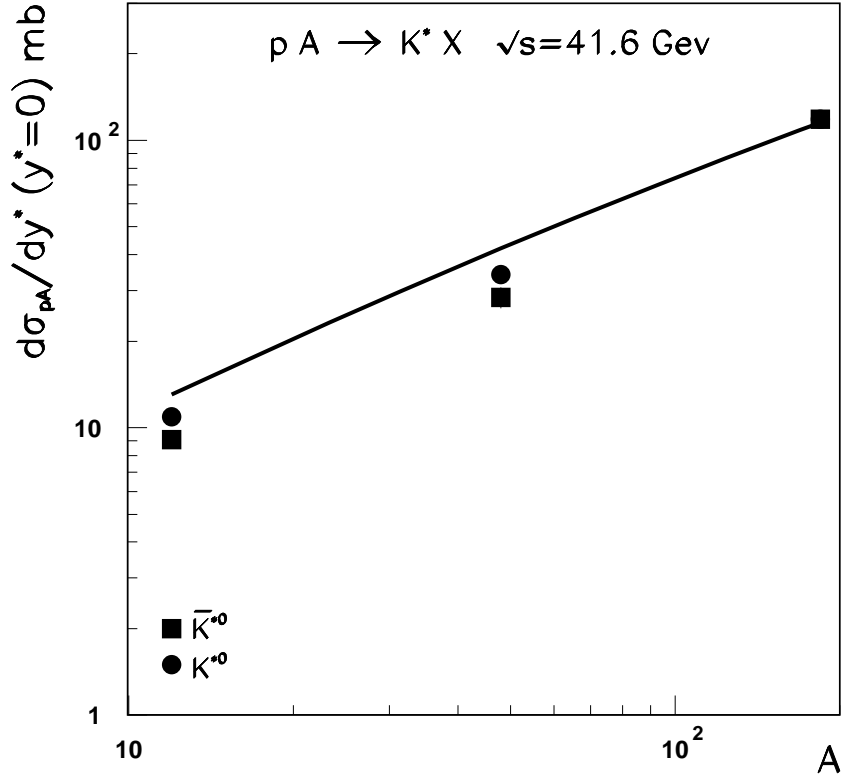}
\vskip -1.0cm
\caption{\footnotesize
Comparison of the results of QGSM calculations for the $A$-dependence of $K^{*0}$ (solid line)
and ${\bar K}^{*0}$ meson production in $pA$ collisions at $\sqrt s = 41.6$~Gev with the corresponding
experimental data by the HERAb Collaboration~\cite{HERAb}. The curve representing the QGSM prediction for
${\bar K}^{*0}$-meson production practically coincides with the solid curve for $K^{*0}$, and so it is not shown.
}
\label{fig:fig8}
\end{figure}

\begin{center}
\vskip -10pt
\begin{tabular}{|c|c|c|c|c|c|} \hline
   &    &  &  &  & \\
Reaction & Produced & Experimental data &  & QGSM & QGSM \\
($\sqrt{s}=41.6\ GeV$) & particle & $\sigma^{prod}_{pA} mb$ & $\sigma_{vis} (mb)$  & total & visible \\
\hline
   &    &  &  &  & \\
$p + C$ & $K^{*0}$ & 43.9 $\pm$ 0.6 $\pm$ 3.3 & 12.1 $\pm$ 0.2 $\pm$ 0.9  & 36.24 & 15.03 \\ \hline
   &    &  &  &  & \\
$p + Ti$ & $K^{*0}$ & 141.2 $\pm$ 2.6 $\pm$ 10.6 & 38.5 $\pm$ 0.7 $\pm$ 2.7 & 124.74 & 46.07 \\ \hline
   &    &  &  &  & \\
$p + W$ & $K^{*0}$ & 465.9 $\pm$ 6.4 $\pm$ 32.7 & 127.5 $\pm$ 1.7 $\pm$ 8.3 & 402.03 & 153.86  \\ \hline
   &    &  &  &  & \\   
$p + C$ & ${\bar K}^{*0}$ & 36.0 $\pm$ 0.6 $\pm$ 2.9 & 10.0 $\pm$ 0.2 $\pm$ 0.7  & 29.04 & 11.25 \\ \hline
   &    &  &  &  & \\
$p + Ti$ & ${\bar K}^{*0}$ & 111.5 $\pm$ 2.5 $\pm$ 9.7 & 31.1 $\pm$ 0.7 $\pm$ 2.3 & 100.03 & 40.34 \\ \hline
   &    &  &  &  & \\
$p + W$ & ${\bar K}^{*0}$ & 388.8 $\pm$ 6.9 $\pm$ 30.8 & 107.6 $\pm$ 1.9 $\pm$ 7.0 & 318.52 & 133.34 \\ \hline
\end{tabular}
\end{center}
\vskip -.1cm
Table~2: {\footnotesize Comparison of the results of the QGSM calculations for $K^{*0}$ and ${\bar K}^{*0}$ meson  
production $\sigma_{prod}$ and for visible $\sigma_{vis}$ cross sections in $pA$ collisions with the corresponding experimental data
by the HERAb Collaboration at $\sqrt{s}=41.6$ GeV~\cite{HERAb}.
} 

\begin{center}
\vskip -12pt
\begin{tabular}{|c|c|c|c|c|} \hline
   &    &  &     &   \\
 Reaction & Centrality & Energy & Experimental data & QGSM \\
	&  & $\sqrt{s}$ $(GeV)$ & $dn/dy$, $|y|\leq 0.5$ &  \\
\hline

   &    &  &     &   \\
$Au + Au \rightarrow$ $K^{*0}$ + X & 0 - 20\% & 62.4 & 6.4 $\pm$ 0.4 $\pm$ 0.7 \cite{kstar2011} & 5.97 \\ \hline
&    &  &  &  \\ 
$Au + Au \rightarrow$  $K^{*0}$ + X & 0 - 5\% & 130.0 & 10.0 $\pm$ 0.9 \cite{kstar130adl} & 9.25 \\ \hline
   &    &  &  & \\ 
$Au + Au \rightarrow$  $K^{*0}$ + X & min. bias & 130.0 & 4.5 $\pm$ 0.7 $\pm$ 1.4 \cite{kstar130adl} & 2.52 \\ \hline
   &    &  &     &    \\ 
$Au + Au \rightarrow$  $K^{*0}$ + X & 0 - 109 & 200.0 & 10.48 $\pm$ 1.45 $\pm$ 1.94 \cite{star200kstar} & 9.83 \\ \hline
   &    &  &     &    \\
$Au + Au \rightarrow$  $K^{*0}$ + X & 0 - 10\% & 200.0 & 9.05 $\pm$ 0.57 $\pm$ 1.01 \cite{kstar2011} & 9.83 \\ \hline
   &    &  &     &    \\
$Cu + Cu \rightarrow$  $K^{*0}$ + X & 0 - 20\% & 200.0 & 2.96 $\pm$ 0.12 $\pm$ 0.3 \cite{kstar2011}  &  2.9 \\ \hline
\end{tabular}
\end{center}
\vskip -.1cm
Table~3: {\footnotesize Experimental data on $dn/dy$, $|y|\leq 0.5$, 
of average neutral $K^{*0}=(K^{*0}+\bar K^{*0})/2$ meson production in different central
nucleus-nucleus collisions at RHIC energies, together with the results of the corresponding QGSM calculations.}

In Table 3 the experimental data on the inclusiive densities $dn/dy$ for average $K^{*0}$=($K^{*0}$ + $\bar K^{*0}$)/2 
mesons produced at RHIC (STAR Collaboration, $\sqrt{s} = 62.4$~Gev~\cite{kstar2011}, $\sqrt{s}=130$~Gev~\cite{kstar130adl},
and $\sqrt{s}=200$~Gev~\cite{star200kstar,kstar2011}) in the midrapidity region for different 
nucleus-nucleus collisions, are compared with the results of the corresponding QGSM calculations.

The QGSM predictions presented in this section show that
the inelastic shadowing effects for neutral vector $K^*$-meson production are very weak at RHIC energies,
and they are not visible inside the experimental error bars.

\section{$K^*$-meson production on nuclear targets at LHC energies}

In this section we will consider the $K^{*0}$ and ${\bar K}^{*0}$-meson production in proton-nucleus and nucleus-nucleus 
collisions at LHC energies.

In the case of production of such a particles as pions and kaons, which give the main contribution to 
mean multiplicity at RHIC energies, the shadowing effects already appear at RHIC~\cite{CKTr}. In the case of
$K^*$-meson production, as well as in the case of $\varphi$-meson production~\cite{amsnphi}, 
the inelastic shadowing effects only appear at LHC energies. 
This behaviour can be connected with the relatively large mass of $K^*$.
The inelastic shadowing effects for pions and kaons were experimentally confirmed by PHoBOS and PHENIX collaborations~\cite{Phob,Phen,MPSd}. 

These saturation effects can be explained by the inelastic screening
corrections due to the multipomeron interactions~\cite{CKTr}. These corrections are negligibly small 
at low energies because of the suppression of the longitudional part of nuclear form factor, but as 
this suppression of the longitudinal part of the nuclear form factor disappears whith the growth of 
the initial energy, the inelastic screening corrections become more and more significant as the 
initial energy increases. 

The calculations of inclusive densities and multiplicities, both in $pA$~\cite{CP1,CP2}, and in heavy ion
collisions~\cite{CP2,CP3}, with accounting for inelastic nuclear screening, can be performed in the 
percolation approach, and they result in a good agreement with the experimental data for a wide energy range.

The percolation approach assumes two or several Pomerons to overlap in 
the transverse space and to fuse in a single Pomeron. Given a certain transverse radius, when the number 
of Pomerons in the interaction region increases, at least part of them may appear inside another Pomeron.
As a result, the internal partons (quarks and gluons) can split, leading to the saturation of the final 
inclusive density. This effect persists with the energy growth until all the Pomerons will overlap~\cite{Dias,Paj,Braun}.

In order to account for the percolation effects in the QGSM, it is technically more simple~\cite{MPSd} to 
consider the maximal number of Pomerons $n_{max}$ emitted by one nucleon in the central region.
After they are cut, these Pomerons lead to the different final states. Then the contributions of all the diagrams
with $n \leq n_{max}$ are inclided into the analysis, as at lower energies. According to the unitarity constraint,
a larger number of Pomerons $n > n_{max}$ can be emitted, but due to fusion in the final state 
(on the quark-gluon string stage) the cut of $n > n_{max}$ Pomerons results in the same final state as 
the cut of $n_{max}$ Pomerons.

Here, $n_{max}$ is a free parameter determined for $K^*$-meson production by comparison with experimental data 
at LHC energies, that it seems can be energy dependent~\cite{BP}.

In Table~4 we present the experimental data by the ALICE Collaboration for the production density 
$dn/dy$ of average $K^{*0}$ =$(K^{*0}$ + $\bar K^{*0})$/2 mesons
in the rapidity range $-0.5 \leq y \leq 0.5$ in $Pb+Pb$ collisions at $\sqrt{s} = 2.76$~TeV,
for centrality 0-5\%~\cite{alicepppb2017} and 0-20\%~\cite{alicepb}, together with the non-single diffraction (NSD) data
for $K^{*0}$-meson production in $p+Pb$ collisions at $\sqrt{s} = 5.02$~TeV~\cite{aliceppb2016},
together with the results of the corresponding QGSM calculations. 

\begin{center}
\vskip -10pt
\begin{tabular}{|c|c|c|c|c|c|} \hline
   &    &  &   &   & \\
Reaction & Centrality & Energy & Experimental data & QGSM & QGSM \\
 &  &  $\sqrt{s}$~(TeV) & $dn/dy$, $|y|\leq 0.5$ & & no screening  \\
\hline
   &    &  &   &   & \\
$Pb + Pb \rightarrow$  $K^{*0}$ + X & 0 - 5\% & 2.76 & 19.56 $\pm$ 2.64~\cite{alicepppb2017}& 20.1 &48.83 \\ \hline
   &    &  &  &   & \\
$Pb + Pb \rightarrow$  $K^{*0}$ + X & 0 - 20\% & 2.76 & 16.6 $\pm$ 2.57~\cite{alicepb}& 16.6 &44.00 \\ \hline
   &    &  &  &   & \\
$p + Pb \rightarrow$  $K^{*0}$ + X & NSD & 5.02 &  0.315 $\pm$ 0.026~\cite{aliceppb2016} & 0.312 & 0.56 \\ \hline
\hline
\end{tabular}
\end{center}
\vskip 0.025cm

\noindent
Table~4: {\footnotesize Experimental data on $dn/dy$, $|y|\leq 0.5$ 
of average neutral $K^{*0}=(K^{*0}+\bar K^{*0})/2 $ meson production in central $Pb+Pb$ and NSD $p+Pb$ 
collisions at LHC energies, together with the results of the corresponding QGSM calculations.} 

For the energy $\sqrt s = 2.76$~ Tev, the model calculation correctly describes the experimental data on density 
$dn/dy$ for average $K^{*0}$ production in $Pb+Pb$ collisions~\cite{alicepppb2017,alicepb} for both centralities 
with a value $n_{max} = 30$. 
The calculations without screening corrections (infinite $n_{max}$), provide values for the density $dn/dy$ significantly
higher than the experimental ones, as it is shown in the last column of Table~4. 

The calculated density $dn/dy$, $-0.5 \leq y \leq 0$ for $p+Pb$ collisions at $\sqrt s = 5.02$~TeV~\cite{aliceppb2016} 
corresponds to $n_{max}$=31. The experimental point on minimum bias spectrum has been normalised to the number of
non-sinle-diffractive (NSD) events. The NSD events include double-diffractive interactions, where both nucleons 
break-up producing particles separated by a large rapidity gap, and other inelastic interactions. 

In the case of $pp$ collisions at $\sqrt{s} = 900$~GeV, the difference between NSD and all inelastic events
is smaller then 15\%, and this difference should be significantly smaller in proton-nucleus collisions. Thus, in the two-channel
model this differense is only about 3\%~\cite{Sh3}. We can then neglect this difference already in proton-nucelus collisions, so
in Table 4 we compare the experinental data on NSD collisions with our calculatins for all inelastic collisions. 

The calculation of the density $dn/dy$ with infinitelyy large $n_{max}$, that corresponds to the absence of shadowing,
gives much higher values for $K^{*0}$ production, both in  in $p+Pb$ collisions, as in $Pb+Pb$ collisions. 

\section{Conclusion}

The QGSM provides a reasonable description of $K^*$-meson production for hadron-proton, proton-nucleus, 
and nucleus-nucleus interactions at not very high (up to RHIC) energies, without adding any new parameters with 
respect to the theoretically based (not free) parameters used to obtain the corresponding description of pion and kaon production. 

The QGSM predictions for the A-dependences of $K^*$-meson production at $\sqrt s = 41$~Gev have the usual behaviour 
$d\sigma/dy (y=0) \propto A^1$, as it is shown in Fig.~8.

In this paper we show that the dependence of $K^*$-meson production on the initial energies in the midrapidity region 
differs from the case of production of light hadrons, as pions and kaons, since for $K^*$-mesons these inelastic screening effects 
are weaker and not visible inside the experimental error bars, up to the RHIC energies.

To include the inelastic shadowing contribution in the analysis of $K^*$-meson production, a new parameter $n_{max}$, 
determined by comparison with experimantal data (see Section 5), is used.

Finally, we show that the inelastic screening effects for the case of $K^*$-meson production only begin to be visible at LHC energies,
where we quantitatively estimate their importance.  

This work has been supported by Russian grant RSGSS-3628.2008.2,
by Ministerio de Ciencia e Innovaci\'on of Spain under project
FPA2017-83814-P, and Maria de Maeztu Unit of Excellence MDM-2016-0692,
and by Xunta de Galicia, Spain, under 2015-AEFIS (2015-PG034), AGRUP2015/11.


\begin{thebibliography}{**}

\bibitem{KTM} A.B. Kaidalov and K.A. Ter-Martorisyan,
Sov. J. Nucl. Phys. {\bf 39}, 979 (1984) and Yad. Fiz. {\bf 39}, 1545 (1984).

\bibitem{K20} A.B. Kaidalov, Phys. At. Nucl. {\bf 66}, 1994 (2003), doi:10.1134/1.1625743,
and Yad. Fiz., {\bf 66}, 2044 (2003).

\bibitem{amsphi}  G.H. Arakelyan, C. Merino, and Yu.M.~Shabelski, Phys. Rev. {\bf D90}, 114019 (2014),
doi:10.1103/PhysRevD.95.074013, and arXiv:1604.01918[hep-ph].

\bibitem{amsnphi}  G.H. Arakelyan, C. Merino, and Yu.M.~Shabelski,  
Phys. At. Nucl. {\bf 80}, 1197 (2017),  10.1134/S1063778817060035; Yad. Fiz. {\bf 80}, 719 (2017); 
and  arXiv:1610.06039[nucl-th].

\bibitem{KTMS} A.B. Kaidalov, K.A. Ter-Martirosyan, and Yu.M.~Shabelski,
Sov. J. Nucl. Phys. {\bf 43}, 822 (1986) and Yad. Fiz. {\bf 43}, 1282 (1986).

\bibitem{Sh1} Yu.M. Shabelski, Z. Phys. {\bf C38}, 569 (1988), doi:10.1007/BF01624362.

\bibitem{Sha} Yu.M. Shabelski, Sov. J. Nucl. Phys. {\bf 50}, 149 (1989)
and Yad. Fiz. {\bf 50}, 239 (1989).

\bibitem{AMPSpl} G.H. Arakelyan, C. Merino, C. Pajares, and Yu.M.~Shabelski,
Phys. At. Nucl. {\bf 76}, 316 (2013), doi:10.1134/S1063778813020026, 
Yad. Fiz. {\bf 76}, 348 (2013), and arXiv:1207.6899[hep-ph].

\bibitem{KaPiZ} A.B. Kaidalov, O.I. Piskunova, Z. Phys. {\bf C30}, 145 (1986), doi:10.1007/BF01560688.

\bibitem{Sh} Yu.M. Shabelski, Sov. J. Nucl. Phys. {\bf 44}, 117 (1986) and Yad. Fiz. {\bf 44}, 186 (1986).

\bibitem{AMPS} G.H. Arakelyan, C. Merino, C. Pajares, and Yu.M.~Shabelski,
Eur. Phys. J. {\bf C54}, 577 (2008), doi:10.1140/epjc/s10052-008-0554-1,
and arXiv:0709.3174[hep-ph].

\bibitem{AMPSk} G.H. Arakelyan, C. Merino, C. Pajares, and Yu.M.~Shabelski,
Eur. Phys. J. {\bf A31}, 519 (2007), doi:10.1140/epja/i2006-10282-6,
and arXiv:0610.264[hep-ph].

\bibitem{aryer} G.H. Arakelyan, Sh.S. Eremian, Phys. At. Nucl. {\bf 58}, 1241 (1995) and
Yad. Fiz. {\bf 58}, 1321 (1995).

\bibitem{APSh} G.H. Arakelyan, C. Pajares, and Yu.M. Shabelski,  
Z. Phys. {\bf C73}, 697 (1997), doi:10.1007/s002880050361, and hep-ph/9602348.

\bibitem{pipkstar} M. Aguilar-Benitez {\it et al.}, Z. Phys. {\bf C44}, 531 (1989),
doi:10.1007/BF01549075.

\bibitem{pipkstar2} N.M. Aghababyan {\it et al.}, Z. Phys. {\bf C46}, 387 (1990), doi:10.1007/BF01621026.

\bibitem{AGIres} G.G. Arakelian, A.A. Grigoryan, N.Ya. Ivanov, and A.B.~Kaidalov,
Z. Phys. {\bf C63}, 137 (1994), doi:10.1007/BF01577553.

\bibitem{CKTr} A. Capella, A. Kaidalov, and J. Tran Thanh Van, Acta Physica Hungarica-Heavy Ion
Phys. {\bf A9}, 169 (1999), and hep-ph/9903244.

\bibitem{amshsv} G.H. Arakelyan, C. Merino, Yu.M.~Shabelski, and A.G. Shuvaev, Phys. Rev. D{\bf 95}, 074013 (2017),
doi:10.1103/PhysRevD.95.074013, and arXiv:1604.01918[hep-ph].

\bibitem{ACKS} G.H. Arakelyan, A. Capella, A.B.~Kaidalov, and Yu.M.~Shabelski, 
Eur. Phys. J. {\bf C26}, 81 (2002), doi:10.1007/s10052-002-0977-z, and arXiv:0103.337[hep-ph].

\bibitem{MPSppb} C. Merino, C. Pajares, and Yu.M.~Shabelski, Eur. Phys. J. {\bf C73}, 2266 (2013).

\bibitem{MPSpa} G.H. Arakelyan, C. Merino, and Yu.M.~Shabelski, 
Eur. Phys. J. {\bf A52}, 9 (2016), doi:10.1140/epja/i2016-16009-2, and arXiv:1509.05218[hep-ph].

\bibitem{AGK} V.A. Abramovsky, V.N. Gribov, and O.V.~Kancheli, Sov. J. Nucl. Phys. {\bf 18}, 308 (1974); 
Yad. Fiz. {\bf 18}, 595 (1973).

\bibitem{Kai} A.B. Kaidalov, Sov. J. Nucl. Phys. {\bf 45}, 902 (1987) and
Yad. Fiz. {\bf 45}, 1452 (1987).

\bibitem{KaPi} A.B. Kaidalov and O.I.~Piskounova, Sov. J. Nucl. Phys. {\bf 41}, 816 (1985) and
Yad. Fiz. {\bf 41}, 1278 (1985).

\bibitem{18} A.A. Grigoryan, N.Ya. Ivanov, Sov. J. Nuc. Phys. {\bf43}, 442 (1986)
and  Yad. Fiz. {\bf43}, 693 (1986).

\bibitem{Shab} Yu.M. Shabelski, Z. Phys. {\bf C57}, 409 (1993), doi:10.1007/BF01474336.

\bibitem{Sh3} Yu.M. Shabelski, Sov. J. Nucl. Phys. {\bf 26}, 573 (1977) 
and Yad. Fiz. {\bf 26}, 1084 (1977); Nucl. Phys. {\bf B132}, 491 (1978), doi:10.1016/0550-3213(78)90473-X.

\bibitem{BT} L. Bertocchi and D. Treleani, J. Phys. {\bf G3}, 147 (1977), doi:10.1088/0305-4616/3/2/007.

\bibitem{Weis} J.H.~ Weis, Acta Phys. Polonica {\bf B7}, 851 (1976).

\bibitem{Jar} T. Jaroszewicz {\it et al.}, Z. Phys. {\bf C1}, 181 (1979), doi:10.1007/BF01445409.

\bibitem{JDDS} J. Dias de Deus and Yu.M. Shabelski, Phys. At. Nucl. {\bf 71}, 190 (2008)
and Yad. Fiz. {\bf 71}, 191 (2008), doi:10.1007/s11450-008-1021-z,10.1134/S1063778808010213,
and hep-ph/0612346.

\bibitem{Alk} G.D. Alkhazov {\it et al.}, Nucl. Phys. {\bf A280}, 330 (1977),
doi:10.1016/0375-9474(77)90609-1.

\bibitem{na49kstar} T. Anticic {\it et al.}, NA49 Collaboration, Phys. Rev. {\bf C84}, 064909 (2011),
doi:10.1103/PhysRevC.84.064909, and arXiv:1105.3109[nucl-ex].

\bibitem{agben}	M.Aguilar-Benitez {\it et al.}, LEBC-EHS Collaboration, Z. Phys. {\bf C50}, 
405 (1991), doi:10.1007/BF01551452.

\bibitem{star200kstar} J.~Adams {\it at al.}, STAR Collaboration, Phys. Rev. {\bf C 71}, 064902 (2005),
doi:10.1103/PhysRevC.71.064902, and nucl-ex/0412019.

\bibitem{alicepppb2017} J. Adam {\it et al.}, ALICE Collaboration, Phys. Rev. {\bf C95}, 064606 (2017),
doi:10.1103/PhysRevC.95.064606, and arXiv:1702.00555[nucl-ex].

\bibitem{alicepp7} B. Abelev {\it et al.}, ALICE Collaboration, Eur. Phys. J. {\bf C72}, 2183 (2012),
doi:10.1140/epjc/s10052-012-2183-y, and arXiv:1208.5717[hep-ex].

\bibitem{HERAb} I. Abt {\it et al.}, HERA-B Collaboration, Eur. Phys. J. {\bf C50}, 315 (2007),
doi:10.1140/epjc/s10052-007-0237-3, and hep-ex/0606049.

\bibitem{kstar2011} M.M. Aggarwal {\it et al.}, STAR Collaboration, Phys. Rev. {\bf C 84}, 034909 (2011),
doi:10.1103/PhysRevC.84.034909, and arXiv:1006.1961[nucl-ex].  

\bibitem{kstar130adl} C.~Adler {\it at al.}, STAR Collaboration, Phys. Rev. {\bf C 66}, 061901 (2002),
doi:10.1103/PhysRevC.66.061901, and nucl-ex/0205015.

\bibitem{Phob} B.B. Back {\it et al.}, PHOBOS Collaboraboration, Phys. Rev. Lett.
{\bf 85}, 3100 (2000), doi:10.1103/PhysRevLett.85.3100, and hep-ex/0007036.

\bibitem{Phen} K.~Adcox {\it at al.}, PHENIX Collaboration, Phys. Rev. Lett. {\bf 86}, 3500 (2001),
doi:10.1103/PhysRevLett.86.3500, and nucl-ex/0012008.

\bibitem{MPSd} C. Merino, C. Pajares, and Yu.M. Shabelski, Eur. Phys. J.
{\bf C59}, 691 (2009), doi:10.1140/epjc/s10052-008-0810-4, and arXiv:0802.2195[hep-ph].

\bibitem{CP1} I. Bautista, C. Pajares, and J. Dias de Deus, Nucl. Phys.
{\bf A882}, 44 (2012), doi:10.1016/j.nuclphysa.2012.03.003, and arXiv:1110.4740[nucl-th].

\bibitem{CP2} I. Bautista, J. Dias de Deus, G. Milhano, and C. Pajares,
Phys. Lett. {\bf B715}, 230 (2012), doi:10.1016/j.physletb.2012.07.029,
and arXiv:1204.1457[nucl-th].

\bibitem{CP3} I. Bautista, C. Pajares, G. Milhano, and  J. Dias de Deus,
Phys. Rev. {\bf C86}, 034909 (2012), doi:10.1103/PhysRevC.86.034909, and
arXiv:1206.6737[nucl-th].

\bibitem{Dias} J. Dias de Deus, E. G. Ferreiro, C. Pajares, and R. Ugoccioni,
Eur. Phys. J. {\bf C40}, 229 (2005), doi:10.1140/epjc/s2005-02127-y,
and hep-ph/0304068.

\bibitem{Paj} C. Pajares, Eur. Phys. J. {\bf C43}, 9 (2005), doi:10.1140/epjc/s2005-02179-y,
and hep-ph/0501125.

\bibitem{Braun} M.A. Braun, E.G. Ferreiro, F. del Moral, and C. Pajares, 
Eur. Phys. J. {\bf C25}, 249 (2002), doi:10.1007/s10052-002-0989-8, and hep-ph/0111378.

\bibitem{BP} M.A. Braun, C. Pajares, Eur. Phys. J. {\bf C16}, 349 (2000), doi:10.1007/s100520050027; 
 M.A. Braun, R.S. Kolevatov, C. Pajares, V.V. Vechernin, Eur. Phys. J. {\bf C32}, 535 (2004), 
DOI: 10.1140/epjc/s2003-01443-6

\bibitem{alicepb} B. Abelev {\it et al.}, ALICE Collaboration, Phys. Rev. {\bf C91}, 
024609 (2015), doi:10.1103/PhysRevC.91.024609, and arXiv:1404.0495[nucl-ex]. 

\bibitem{aliceppb2016} J. Adam {\it et al.}, ALICE Collaboration, Eur. Phys. J. {\bf C76}, 245 (2016),
doi:10.1140/epjc/s10052-016-4088-7, and arXiv:1601.07868.

\end{thebibliography}
\end{document}